# Nonlinear absorption study in four and five energy level systems

A Srinivasa Rao

Department of Physics, Pondicherry University, Puducherry, India-605014

Present address: Photonic Sciences Lab, Physical Research Laboratory, Ahmedabad, India-380009

Email: sri.jsp7@gmail.com, asvrao@gmail.com

**Abstract**

Nonlinear absorption in four and five energy level systems have been studied with the aid of steady-state rate equation approach. We report on the tunability of saturable and reverse saturable absorption as a function of spectroscopic parameters (lifetimes and absorption cross-sections). Detailed information is given on the estimation of spectroscopic parameters in the nonlinear absorption spectroscopy. The exhaustive graphical analysis of this article can provide the brief idea about transmittance curves of the nonlinear absorption to the experimentalists. In four and five level cascade models, simultaneously saturable and reverse saturable absorption can be generated.

**1 Introduction**

In past two decades, nonlinear absorption attached a great interest in the optical field due to its wide range of applications in the spectroscopic study. Rate equation technique is an accomplished method in the nonlinear absorption spectroscopy. The spectroscopic parameters (lifetimes and absorption cross-sections) can be measured through pump-probe absorption spectroscopy [1, 2], single beam transmittance (SBT) [3-5] and z-scan [6-8]. while, pump-probe deals with non-degenerate nonlinear absorption spectroscopy, z-scan and SBT are used in degenerate nonlinear absorption spectroscopy. Among these techniques, z-scan is a very simplest one in the spectroscopic study. However, it is insufficient, when it comes to the more than three energy levels contributed in the absorption process due to its finite longitudinal intensity profile restricted by focusing lens. Moreover in the z-scan, the sample has to cross the focal plane in the absorption study and to avoid the sample damage, while it is crossing the focus, we have to focus very low power and which is insufficient to the higher excited state study. But through SBT one can overcome this flaw by its extended tunable range which is determined by laser power. In SBT, the available high power lasers fulfills this criterion in the absence focused laser beam.

Experimentally researchers have seen the four and five energy levels participation in the degenerate absorption process [9-20]. But still, no article has discussed absorption cross-sections and lifetimes effect in the nonlinear absorption process. In this article we have shown, how one can see the single saturable absorption (SSA) [21], double saturable absorption (DSA) [21] and saturable with reverse saturable absorption (SRSA). In addition to this, we systematically study the influence of spectroscopic parameters in the transmittance curve through SBT. Following parameters are considered for simulations in the throughout the paper: excitation wavelength ($\lambda$) is 532 nm, interacting length of the material with the laser beam ($L$) is 0.1mm and number density of absorption species (N) is $10^{19}$ absorption species/cm$^3$. The number density of molecules in each state $|i\rangle$ is taken to be $N_i$ with satisfying the condition $N=\Sigma N_i$ and the fractional number density in each state is $n_i=N_i/N$ and $\Sigma n_i=1$. The pumping rate from $|i\rangle$ to $|j\rangle$ level is $W_{ij}=(I/h\nu)\sigma_{ij}$, where $\sigma_{ij}$ is the absorption cross-section from $|i\rangle$ to $|j\rangle$ level. The decay time from $|j\rangle$ to $|i\rangle$ level is $\tau_{ji}$. In all the models, the rate equations has been solved by considering steady state population i.e., $dn_i/dt=0$. The transmittance of the interacting material is given by T=exp($-\alpha_{eff}L$), where $\alpha_{eff}$ is effective absorption coefficient, which is differer for each energy model.

**2 Four level cascade system**

In four level cascade systems, subsequent energy levels are separated by equal energies ($E_j-E_i$=photon energy, $j=i+1$ ) as depicted in figure 1.

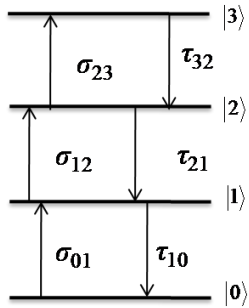

Figure 1. Four level cascade system

Absorption takes place through three stimulated absorption cross-sections ($\sigma_{ij}$, $j=i+1$) in the cascade way and the downward transitions carried through stimulated emission cross-sections ($\sigma_{ji}$, $j=i+1$) and decay time of energy levels ($\tau_{ji}$, $j=i+1$). According to Einstein stimulated excitation theory, stimulated absorption cross-section is equal to the stimulated emission cross-section i.e., $W_{ij}=W_{ji}$ [22]. Thus the rate equations have the form as

$$n_0 + n_1 + n_2 + n_3 = 1 \tag{1a}$$

$$\frac{dn_0}{dt} = -W_{01}(n_0 - n_1) + \frac{n_1}{\tau_{10}} \tag{1b}$$

$$\frac{dn_1}{dt} = W_{01}(n_0 - n_1) - W_{12}(n_1 - n_2) - \frac{n_1}{\tau_{10}} + \frac{n_2}{\tau_{21}} \tag{1c}$$

$$\frac{dn_2}{dt} = W_{12}(n_1 - n_2) - W_{23}(n_2 - n_3) - \frac{n_2}{\tau_{21}} + \frac{n_3}{\tau_{32}} \tag{1d}$$

$$\frac{dn_3}{dt} = W_{23}(n_2 - n_3) - \frac{n_3}{\tau_{32}} \tag{1e}$$

The rate equations 1(a)-1(e) are numerically solved and obtained the population in each energy level [21]. The absorption coefficient is $\alpha_{eff}=\alpha_0(n_0-n_1)+\alpha_1(n_1-n_2)+\alpha_2(n_2-n_3)$. Here, $\alpha_0=N\sigma_{01}$, $\alpha_1=N\sigma_{12}$, and $\alpha_2=N\sigma_{23}$ are respective linear absorption coefficients from $|0\rangle$, $|1\rangle$, and $|2\rangle$ energy levels to their next higher energy levels.

The lifetime's effect on the transmittance curve is shown in figure 2. In four level absorption process, the transmittance get saturates two times in the entire transmittance profile and the transmittance have the steps in the absorption processes as: linear absorption/transmittance followed by first saturation in the absorption/transmittance followed by second saturation in the absorption/transmittance [21]. As depicted in figure 2(a), first saturation takes place in the $|1\rangle$ level and the saturation point determined by the decay time $\tau_{10}$. The saturation intensity of the $|1\rangle$ state is given by $I_{S01}=h\nu/\tau_{10}\sigma_{01}$ W/cm$^2$ [4]. The onset of saturation can be tuned by $\tau_{10}$. The second saturation in the transmittance curve is controlled by $I_{S23}=h\nu/\tau_{32}\sigma_{23}$ W/cm$^2$ as illustrated in figure 2(c). The onset of the second saturation can be increased by decreasing the lifetime $\tau_{32}$. The decrease in the lifetime $\tau_{32}$ delays the saturation in the state $|3\rangle$ due to fast transitions from state $|3\rangle$ to the state $|2\rangle$. The saturation intensity of the second state $I_{S12}=h\nu/\tau_{21}\sigma_{12}$ couples the two saturation intensities $I_{S01}$ and $I_{S23}$. Clearly, we can see the following three process in figure 2(b). when $\tau_{21}=10^{-8}$ s, the second saturation intensity $I_{S12}$ ($2h\nu \times 10^{25}$ W/cm$^2$) is close to the first saturation intensity $I_{S01}$ ($h\nu \times 10^{25}$ W/cm$^2$) and the two states acquire the saturation almost at the same intensity. At $\tau_{21}=10^{-9}$ s, the second saturation intensity $I_{S12}$ ($2h\nu \times 10^{26}$ W/cm$^2$) is larger than the $I_{S01}$ ($h\nu \times 10^{25}$ W/cm$^2$) and the transmittance curve have two saturation intensities slightly at two different positions. As a result, we can see that the two saturation intensities $I_{S01}$ and $I_{S23}$ are smoothly connected by $I_{S12}$. $\tau_{21}=10^{-10}$ s shows clear double saturable absorption (DSA) and corresponding saturation intensity $I_{S12} = 2h\nu \times 10^{27}$ W/cm$^2$. The transmittance curve is more sensitive to the lower energy levels lifetime as compared with the higher energy levels because the population in the higher energy levels always less than the lower energy levels even we excite with ultra-short pulses. This behavior can be seen in figure 2. Each of three lifetimes controlling the transmittance at different parts of the transmittance curve, so experimentally one can obtain the three lifetimes in the SBT. As explained about the sensitivity of lifetimes in the transmittance curve, the accuracy of the estimation of lifetimes from transmittance curve follows accuracy as accuracy ($\tau_{10}$)> accuracy ($\tau_{21}$)> accuracy ($\tau_{32}$).

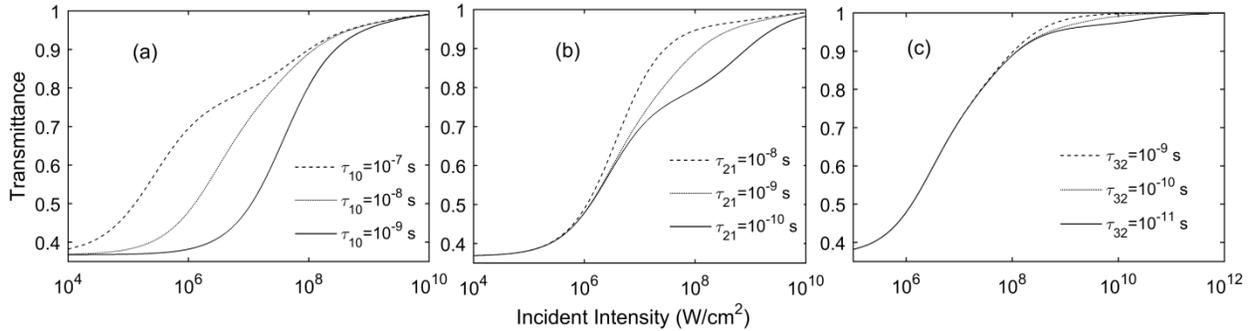

Figure 2. Transmittance curves as a function of lifetimes: $\tau_{10}=10^{-8}$ s (variable only in figure a), $\tau_{21}=10^{-9}$ s (variable only in figure b), $\tau_{32}=10^{-10}$ s (variable in only figure c), $\sigma_{01}=10^{-17}$ cm$^2$, $\sigma_{12}=5\times10^{-18}$ cm$^2$, and $\sigma_{23}=10^{-18}$ cm$^2$.

The transmittance curve as a function of absorption cross-sections illustrated in figure 3. While the onset of saturation depends on the lifetimes (figure 2), the nature of the absorption depends on the absorption cross-sections (figure 3). We can see in figure 3(a), the transmittance of the linear absorption can be tuned by ground state absorption cross-section ($\sigma_{01}$). Four level system at first saturation have the SA nature, if $\sigma_{01}$ ($x\times10^{-18}$ cm$^2$, $x$>5) > $\sigma_{12}$ ($5\times10^{-18}$ cm$^2$) and have the RSA character when $\sigma_{01}$ ($1\times10^{-18}$ cm$^2$) < $\sigma_{12}$ ($5\times10^{-18}$ cm$^2$). Figure 3(b) shows how the $\sigma_{12}$ cross-section modulates the first saturation in the transmittance curve. By tuning $\sigma_{12}$, one can regulate the transmittance curve between SA and RSA with satisfying above discussed conditions. As depicted in figure 3(c), the second saturation of the transmittance curve can be regulated by varying the absorption cross-section $\sigma_{23}$. SSA, DSA, and SRSA can be seen at respective $\sigma_{23} = 5\times10^{-17}$ cm$^2$, $\sigma_{23} = 5\times10^{-18}$ cm$^2$, and $\sigma_{23} = 10^{-18}$ cm$^2$. In the figure 3(c), $\sigma_{23}=5\times10^{-17}$ cm$^2$ cross-section transmittance curve have the condition $\sigma_{01}$ ($10^{-17}$ cm$^2$) > $\sigma_{12}$ ($5\times10^{-18}$ cm$^2$) < $\sigma_{23}$ ($5\times10^{-17}$ cm$^2$). As a consequence, the transmittance curve first saturates and then reverse saturates. While increasing the $\sigma_{23}$ value from $5\times10^{-17}$ cm$^2$ to $1\times10^{-18}$ cm$^2$, we can see that the transmittance curve changing SRSA to DSA and DSA to SSA due to the decrease in the absorption cross-section $\sigma_{23}$. As like lifetimes, the three absorption cross-sections modulates the transmittance curve at different positions. Thus, experimentally we can obtain the cross-sections from the transmittance curve.

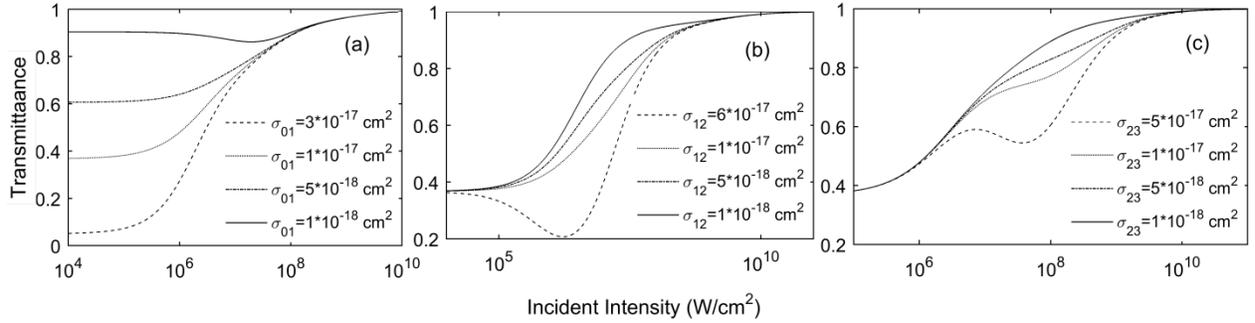

Figure 3. Transmittance curves as a function of absorption cross-sections: $\tau_{10}=10^{-8}$ s, $\tau_{21}=10^{-9}$ s, $\tau_{32}=10^{-10}$ s, $\sigma_{01}=10^{-17}$ cm$^2$ (variable only in figure a), $\sigma_{12}=5\times10^{-18}$ cm$^2$ (variable only in figure b), and $\sigma_{23}=10^{-18}$ cm$^2$ (variable only in figure c).

## 3 Five level cascade system

As like in the four level cascade system, in the five level cascade system, five subsequent energy levels separated by equal energies as shown in figure 4.

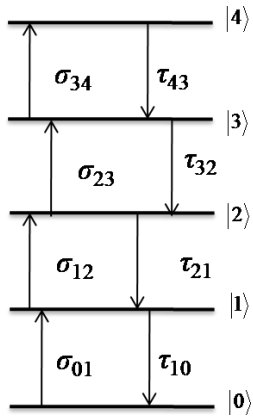

Figure 4. Five level cascade system

The population dynamics among the five energy levels can be understand through rate equations as

$$n_0 + n_1 + n_2 + n_3 + n_4 = 1 \quad (2a)$$

$$\frac{dn_0}{dt} = -W_{01}(n_0 - n_1) + \frac{n_1}{\tau_{10}} \quad (2b)$$

$$\frac{dn_1}{dt} = W_{01}(n_0 - n_1) - W_{12}(n_1 - n_2) - \frac{n_1}{\tau_{10}} + \frac{n_2}{\tau_{21}} \quad (2c)$$

$$\frac{dn_2}{dt} = W_{12}(n_1 - n_2) - W_{23}(n_2 - n_3) - \frac{n_2}{\tau_{21}} + \frac{n_3}{\tau_{32}} \quad (2d)$$

$$\frac{dn_3}{dt} = W_{23}(n_2 - n_3) - W_{34}(n_3 - n_4) - \frac{n_3}{\tau_{32}} + \frac{n_4}{\tau_{43}} \quad (2e)$$

$$\frac{dn_4}{dt} = W_{34}(n_3 - n_4) - \frac{n_4}{\tau_{43}} \quad (2f)$$

The numerical solutions of the rate equations 2(a)-2(f) are used to obtain the absorption coefficient $\alpha_{eff}=\alpha_0(n_0-n_1)+\alpha_1(n_1-n_2)+\alpha_2(n_2-n_3)+\alpha_3(n_3-n_4)$. Here, $\alpha_0=N\sigma_{01}$, $\alpha_1=N\sigma_{12}$, $\alpha_2=N\sigma_{23}$, and $\alpha_3=N\sigma_{34}$ are respective linear absorption coefficients from $|0\rangle$,

$|1\rangle$, $|2\rangle$, and $|3\rangle$ energy levels to their next higher energy level.

Even though four and five level cascade systems looks like similar in the absorption process but the transmittance curves of two models are different in their behavior at high intensity, where fifth energy level contributes. In five level cascade systems, the fifth level generates one more RSA/SA to the transmittance curve discussed in section 2. The graphical study of the transmittance curve as a function of lifetimes is presented in figure 5. We can conclude from figure 5 (a-c) and figure 2(a-c), the effect of $\tau_{10}$, $\tau_{21}$, and $\tau_{32}$ on the transmittance curve in the four and five level models are same. Not only to tune the onset of saturation process by lifetimes but also we can modulate the absorption process as a function of lifetimes. Higher excited states stimulation absorption process determined by the corresponding lifetime of the excited states because the population required for stimulated absorption process in the excited states depends on the lifetime of the corresponding states. We can see following different saturation processes in single transmittance curve: DSA with RSA can be observed at $\tau_{10}=10^{-7}$ s (figure 5(a)). First DSA generated between $|0\rangle$, $|1\rangle$, and $|2\rangle$ states because first saturation ($I_{S01}$) moved towards linear region and away from second saturation ($I_{S12}$) at $\tau_{10}=10^{-7}$ s i.e., $I_{S01} = h\nu \times 10^{24}$ W/cm$^2$ < $I_{S12} = 2h\nu \times 10^{26}$ W/cm$^2$ and with satisfying the condition $\sigma_{01}=10^{-17}$ cm$^2$ > $\sigma_{12}=5\times10^{-18}$ cm$^2$ > $\sigma_{23}=10^{-18}$ cm$^2$ and then RSA generated between $|3\rangle$, and $|4\rangle$ states due to $\sigma_{23}=10^{-18}$ cm$^2$ < $\sigma_{43}=10^{-17}$ cm$^2$. SSA with RSA created in the transmittance curve for $\tau_{10}=10^{-9}$ s & $10^{-8}$ s (figure 5(a)), $\tau_{21}=10^{-8}$ s & $\tau_{21}=10^{-9}$ s (figure 5(b)), $\tau_{32}=10^{-9}$ s & $10^{-10}$ s (figure 5(c)), and $\tau_{43}=10^{-11}$ s & $10^{-12}$ s (figure 5(d)). In this case, $I_{S01}$ and $I_{S12}$ are close to each other and there will not be any population accumulation in the $|1\rangle$ state as a result single saturation generated in the $|1\rangle$, and $|2\rangle$ states. Next RSA generated because of $\sigma_{23}=10^{-18}$ cm$^2$ < $\sigma_{43}=10^{-17}$ cm$^2$. DSA at $\tau_{21}=10^{-10}$ s (figure 5(b)) is observed due to smaller second excited state lifetime ($\tau_{21}$) even though $\sigma_{23}=10^{-18}$ cm$^2$ < $\sigma_{43}=10^{-17}$ cm$^2$ because of less population transfer to the higher excited states ($|3\rangle$, and $|4\rangle$ states). SSA at $\tau_{32}=10^{-11}$ (figure 5(c)) and $\tau_{43}=10^{-10}$ s (figure 5(d)) observed because due to the very short lifetime of the upper excited states. As shown in figure 5(d), One can control the end reverse saturation point of transmittance curve by tuning the lifetime $\tau_{43}$. The four lifetimes action on the transmittance curve does not concur so with the transmittance curve we can measure all the four lifetimes.

As presented in figure 6, the effect of the first three absorption cross-sections $\sigma_{01}$, $\sigma_{12}$, and $\sigma_{23}$ on the transmittance curve are same as like four level cascade systems. These three parameters effect on transmittance curve can be seen in figure 5(a-c). In figure 5(b), at $\sigma_{12}=5\times10^{-17}$ cm$^2$ clearly we can see two RSAs with separated by SA due to large absorption cross-sections of first and third excited states as compared with respective ground and second excited states. Using four absorption cross-sections we can modulate transmittance curve as different combination of SSA, DSA, and RSA. The absorption cross-sections and lifetimes are can be measured independently through SBT process.

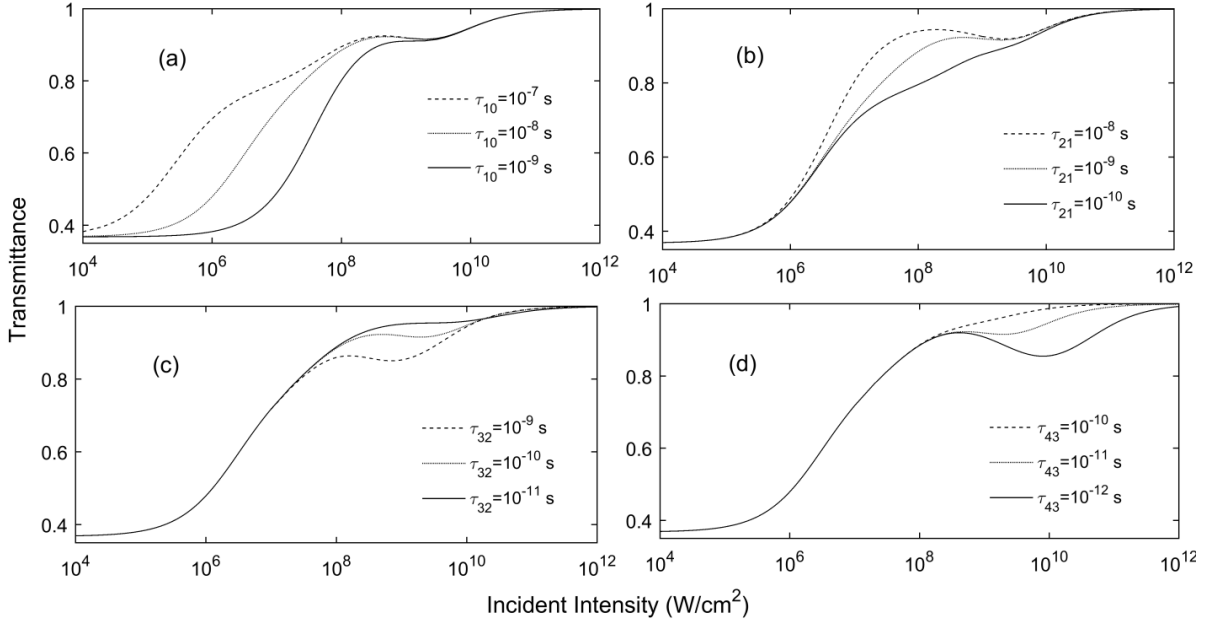

Figure 5. Transmittance curves as a function of lifetimes: $\tau_{10}=10^{-8}$ s (variable only in figure a), $\tau_{21}=10^{-9}$ s (variable only in figure b), $\tau_{32}=10^{-10}$ s (variable only in figure c), $\tau_{43}=10^{-11}$ s (variable only in figure d), $\sigma_{01}=10^{-17}$ cm$^2$, $\sigma_{12}=5\times10^{-18}$ cm$^2$, $\sigma_{23}=10^{-18}$ cm$^2$ and $\sigma_{34}=10^{-17}$ cm$^2$.

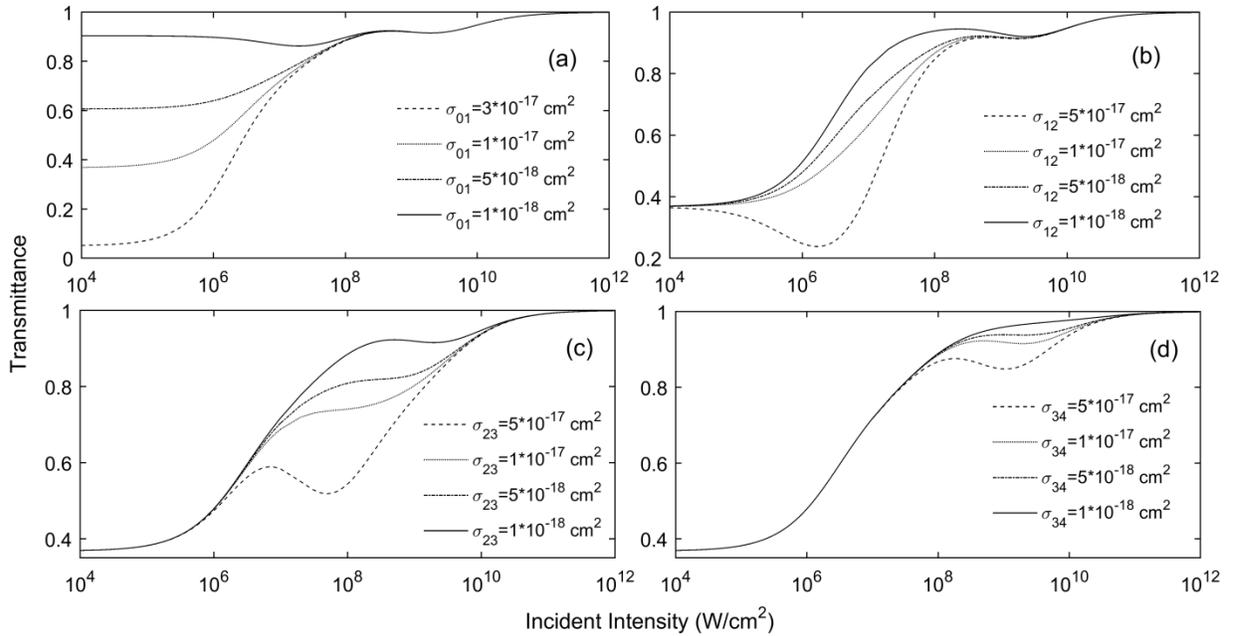

Figure 6. Transmittance curves as a function of absorption cross-sections: $\tau_{10}=10^{-8}$ s, $\tau_{21}=10^{-9}$ s, $\tau_{32}=10^{-10}$ s, $\tau_{43}=10^{-11}$ s, $\sigma_{01}=10^{-17}$ cm$^2$ (variable only in figure a), $\sigma_{12}=5\times10^{-18}$ cm$^2$ (variable only in figure b), $\sigma_{23}=10^{-18}$ cm$^2$ (variable only in figure c) and $\sigma_{34}=10^{-17}$ cm$^2$. (variable only in figure d).

## 4 Five level non-cascade system

Four level non-cascade system is discussed in the Ref. [21]. In this section, we will discuss five level non-cascade system. As depicted in figure 7, in five level non cascade systems three singlet states ($|0\rangle$, $|2\rangle$, and $|4\rangle$) and two triplet states ($|1\rangle$ and $|3\rangle$) participates in the absorption process. singlet and triplet states are spin coupled and the transitions from singlet to triplet states take place through non-radiative decay. The best example of singlet-triplet state molecules are organic and metal−organic materials are of interest for many applications including optical limiting [23-25].

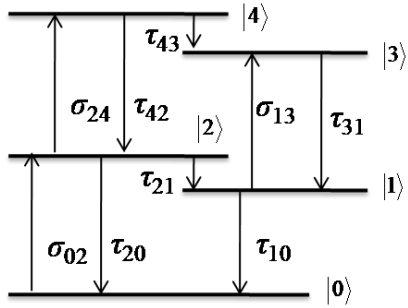

Figure 7. Five level non-cascade system

The population dynamics among the five levels in the differential equations form as given in equation 3.

$$n_0 + n_1 + n_2 + n_3 + n_4 = 1 \quad (3a)$$

$$\frac{dn_0}{dt} = -W_{02}(n_0 - n_2) + \frac{n_1}{\tau_{10}} + \frac{n_2}{\tau_{20}} \quad (3b)$$

$$\frac{dn_1}{dt} = -W_{13}(n_1 - n_3) - \frac{n_1}{\tau_{10}} + \frac{n_2}{\tau_{21}} + \frac{n_3}{\tau_{31}} \quad (3c)$$

$$\frac{dn_2}{dt} = W_{02}(n_0 - n_2) - W_{24}(n_2 - n_4) - \frac{n_2}{\tau_{20}} - \frac{n_2}{\tau_{21}} + \frac{n_4}{\tau_{42}} \quad (3d)$$

$$\frac{dn_3}{dt} = W_{13}(n_1 - n_3) - \frac{n_3}{\tau_{31}} + \frac{n_4}{\tau_{43}} \quad (3e)$$

$$\frac{dn_4}{dt} = W_{24}(n_2 - n_4) - \frac{n_4}{\tau_{43}} - \frac{n_4}{\tau_{42}} \quad (3f)$$

The obtained populations in the energy levels from the numerical simulations of rate equations 3(a)-3(f) are used to obtain the absorption coefficient $\alpha_{eff}=\alpha_0(n_0-n_2)+\alpha_1(n_1-n_3)+\alpha_2(n_2-n_4)$. Here, $\alpha_0=N\sigma_{02}$, $\alpha_1=N\sigma_{13}$, and $\alpha_2=N\sigma_{24}$ are respective linear absorption coefficients from $|0\rangle$, $|1\rangle$, and $|2\rangle$ energy levels to $|2\rangle$, $|3\rangle$, and $|4\rangle$ energy levels.

In non-cascade systems, due to metastable nature of triplet states ($|1\rangle$ and $|3\rangle$), maximum population distributes in the ground and triplet states. Therefore, the major contribution in the absorption process takes place by triplet lifetimes $\tau_{10}$ and $\tau_{31}$ (figure 8(a) and 8(d)). The contribution from the singlet states not so great due to less population in the $|2\rangle$ state. Here, $\tau_{42}$ and $\tau_{43}$ are not discussed because there is no effect from them on the transmittance curve. Due to trapping of atoms/molecules in the triplet states, there will not be sufficient atoms/molecules present in the $|2\rangle$ state to pump the atoms to $|4\rangle$ state. Thus there is no contribution of $|4\rangle$ state on the transmittance curve. From this, we can infer that using CW and nano-second lasers it is possible to measure the triplet states and first excited singlet state lifetimes. Low reputation rate (~1 kHz) ultra-fast laser pulses can be used to study the all the singlet states lifetime. Because of low repetition rate and short pulse nature, the contribution of triplet states on the ultra short-pulses can be successfully neglected [22]. Under low repetition rate ultra short-laser excitation, non-cascade system transforms to cascade system and the transmittance profile becomes as like we discussed in the sections three and four. To complete the spectroscopic study of non cascade system, it is necessary to use either CW or nano-second laser with low repetition rate ultra-short laser pulses.

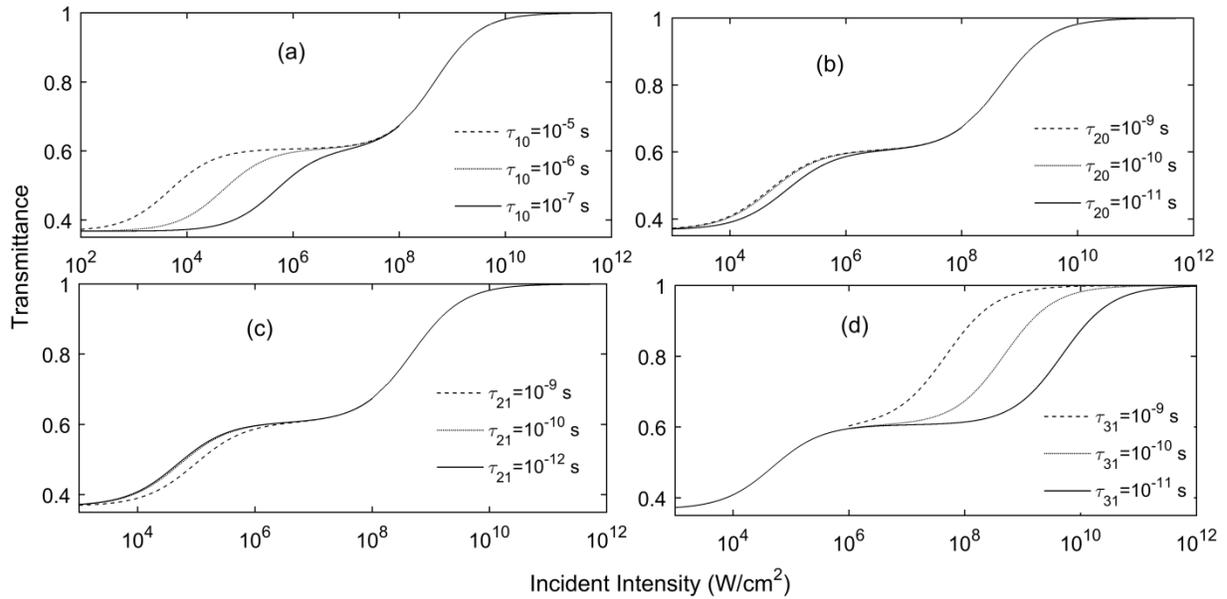

Figure 8. Transmittance curves as a function of lifetimes: $\tau_{10}=10^{-6}$ s (variable only in figure a), $\tau_{20}=10^{-9}$ s (variable only in figure b), $\tau_{21}=10^{-11}$ s (variable only in figure c), $\tau_{31}=10^{-10}$ s (variable only in figure d), $\tau_{42}=10^{-10}$ s, $\tau_{43}=10^{-11}$ s, $\sigma_{02}=10^{-17}$ cm$^2$, $\sigma_{13}=10^{-18}$ cm$^2$, and $\sigma_{24}=10^{-18}$ cm$^2$.

The absorption cross-sections effect on the transmittance curve presented in figure 9. In five level cascade system, two absorption cross sections present and for this reason only SSA and RSA can be generated in the nonlinear absorption process. Linear transmittance value can be tuned with $\sigma_{02}$ and saturation nature can be tuned with $\sigma_{13}$. As explained in the before, due to the minute population in the $|2\rangle$ state, the effect of $\sigma_{24}$ on the transmittance curve negligible and it can be measured by only in the presence of ultra-short laser excitation.

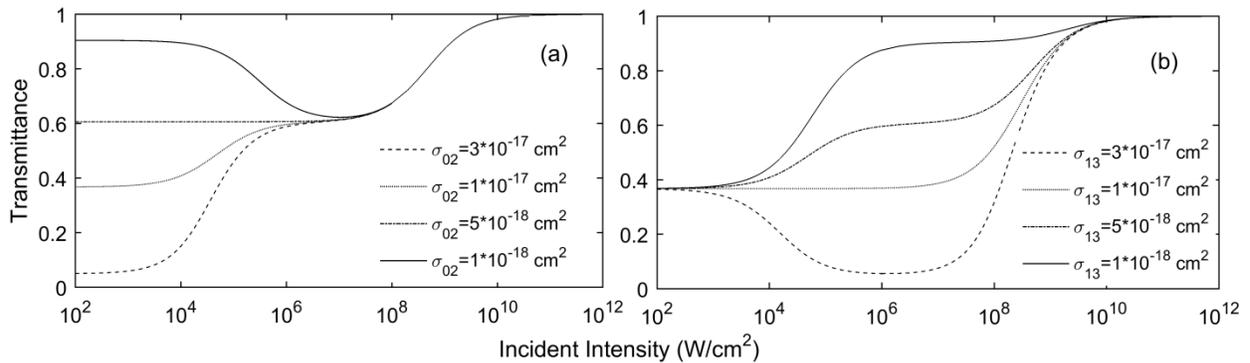

Figure 9. Transmittance curves as a function of absorption cross-sections: $\tau_{10}=10^{-6}$ s, $\tau_{20}=10^{-9}$ s, $\tau_{21}=10^{-11}$ s, $\tau_{31}=10^{-10}$ s, $\tau_{42}=10^{-10}$ s, $\tau_{43}=10^{-11}$ s, $\sigma_{02}=10^{-17}$ cm$^2$, $\sigma_{13}=10^{-18}$ cm$^2$ (variable only in figure a), and $\sigma_{24}=10^{-18}$ cm$^2$ (variable only in figure b).

## 5 Conclusion

Simultaneously saturable and reverse saturable absorption can be generated in the materials those act as four and five level cascade systems. If we use four and five level cascade systems as saturable absorbers for applications then the damage of optical components from the sudden laser fluctuations can be controlled. Because at ambient laser intensity, these systems act as saturable absorbers and in case of any drastic increase in the laser intensity, these systems become reverse saturable absorbers and protect the devices from laser damage by absorbing the unwanted laser intensity. In five level cascade system, we can generate two RSA separated by SA in the transmittance profile. We have shown how to measure the spectroscopic properties of four and five level cascade systems in the SBT. The spectroscopic properties of five level non-cascade systems can be measured by independently exciting the material with CW or nano-second laser and low repetition rate ultra-short laser pulses in the SBT. As explained theoretically in this paper, experimentally one can modulate the transmittance curve by using optical materials in the presence of different dopants, hosts and in different states [23-26].

## References

1. Woutersen A., Emmerichs U., Bakker H.J., Femtosecond mid-IR pump-probe spectroscopy of liquid water: Evidence for a two-component structure, Science 278 (1997) 658-660.

2. Alfano J.C., Walhout P.K., Kimura Y., Barbara P.F., Ultrafast transient-absorption spectroscopy of the aqueous solvated electron, J. Chem. Phys. 98 (1993) 5996-5998.

3. Hercher M., An analysis of saturable absorbers, Appl. Opt. 6 (1967) 947-954.

4. Rao A. S., Comparison of Rate Equation models for Nonlinear Absorption, Optik 158 (2018) 652–663.

5. Rao S., Optical limiting in the presence of simultaneous one and two photon absorption, Optik 157 (2018) 900-905.

6. Vijayakumar S., Adithya A., Sharafudeen K.N., Balakrishna K., Chandrasekharan K., Third-order nonlinear optical properties in 4-[(E)-(2-phenylhydrazinylidene) methyl] tetrazolo [1, 5-a] quinoline doped PMMA thin film using Z-scan technique, J. Mod. Opt. 57 (2010) 670-676.

7. Wu F., Tian, W., Chen W., Zhang, G., Zhao G., Cao S., Xie W., Optical nonlinearity and optical limiting of CdSeS/ZnS quantum dots, J. Mod. Opt. 56 (2009) 1868-1873.

8. Rao A. S., Dar M. H., Venkatramaiah N, Venkatesan R, Sharan A., Third order optical nonlinear studies and its use to estimate thickness of sandwiched films of tetra-phenyl porphyrin derivatives, J. Non. Opt. Phys. Mat. 25 (2016) 1650039.

9. Ji, W., Du H.J., Tang S. H., Shi S., Nanosecond reverse saturable absorption in cubanelike transition-metal clusters, JOSA B 12 (1995) 876-881.

10. Srinivas N.K., Rao S. V., Rao D. V. G. L. N., Kimball B. K., Nakashima M., Decristofano B. S., and Rao D. N., Wavelength dependent studies of nonlinear absorption in zinc meso-tetra (p-methoxyphenyl) tetrabenzoporphyrin (Znmp

TBP) using Z-scan technique, J. Porphy. Phthalocy. 5 (2001) 549-554.

11. Dmitriy I. Kovsh, Sidney Y., David J. H., Eric W. Van S., Nonlinear optical beam propagation for optical limiting, Appl. Opt. 38 (1999) 5168-5180.

12. Divakara Rao K., Anantha Ramakrishna S., Gupta P.K., Nonlinear optical studies in tetraphenyl-porphyrin-doped boric acid glass using picosecond pulses, Appl. Phys. B 72 (2001) 215–219.

13. Zhenrong S., Minghong T., Heping Z., Liangen D., Zugeng W., Jie D., Guoqing B., Zhizhan X., Nanosecond reverse saturable absorption and optical limiting in (Me4N)2[Cd(dmit)(Sph)2] JOSA B 18 (2001) 1464-1468.

14. Sreekumar G., Louie Frobel P G., Muneera C I., athiyamoorthy K., Vijayan C., Chandrachur M., Saturable and reverse saturable absorption and nonlinear refraction in nanoclustered Amido Black dye–polymer films under low power continuous wave He–Ne laser light excitation, J. Opt. A: Pure Appl. Opt. 11 (2009) 125204.

15. Prem Kiran P., Raghunath Reddy D., Bhaskar G. M., Aditya K. D., Ravindra Kumar G., Narayana Rao D., Nonlinear absorption properties of axial-bonding_ type tin(IV) tetratolylporphyrin based hybrid porphyrin arrays, Opt. Commun. 252 (2005) 150–161.

16. Alan K., Lee T., Marvin B. K., Kirk Dougherty T., William E. E., Optical limiting with $C_{60}$ in polymethyl methacrylate, Opt. Lett. 18 (1993) 334-336.

17. Tutt L. W., McCahon S. W., Reverse saturable absorption in metal cluster compounds, Opt. Lett. 15 (1990) 700-702.

18. Santhi A., Namboodiri Vinu V., Radhakrishnan P., Nampoori V. P., Simultaneous determination of nonlinear optical and thermo-optic parameters of liquid samples, Appl. Phys. Lett. 89 (2006) 231113.

19. Venugopal Rao S., Narayana Rao D., Akkara J. A., DeCristofano B.S., Rao D.V.G.L.N., Dispersion studies of non-linear absorption in C using Z-scan, Chem. Phys. Lett. 297 (1998) 491–498.

20. Goncalves P. J., De Boni L., Borissevitch I. E., Zı´lio S. C., Excited State Dynamics of meso-Tetra(sulphonatophenyl) Metalloporphyrins, J. Phys. Chem. A. 112 (2008) 6522–6526.

21. Allam S.R., Theoretical study on nonlinear properties of four level systems under nano-second illumination, Laser Phys. 25 (2015) 055701.

22. Allam S. R., Sharan A., One, two and three photon absorption of two level system in femto-second laser excitation, J. Opt. (2017) 1-6.

23. Hollins R. C., Materials for optical limiters. Current opinion in solid state and materials science. 30 (1999) 189-196.

24. Hughes S., Spruce G., Wherrett B. S., Kobayashi T., Comparison between the optical limiting behavior of chloroaluminum phthalocyanine and a cyanine dye, J. Appl. Phys. 81 (1997) 5905-5912.

25. Sanghadasa M., Shin I. S., Clark R. D., Guo, H., Penn B. G., Optical limiting behavior of octa-decyloxy metallo-phthalocyanines, J. Appl. Phys. 90 (2001) 31-37.

26. Perry J.W., Alvarez D., Choong I., Mansour K., Marder S. R., Perry K. J., Enhanced reverse saturable absorption and optical limiting in heavy-atom-substituted phthalocyanines, Opt. Lett. 19 (1994) 625-627.

27. Ji W., Du H. J., Tang S. H., Shi S., Nanosecond reverse saturable absorption in cubanelike transition-metal clusters, JOSA B 12(1995) 876-881.

28. Tutt L. W., McCahon S. W., Reverse saturable absorption in metal cluster compounds, Opt. Lett. 15(1990) 700-702.

29. Sendhil K., Vijayan C., Kothiyal M. P., Nonlinear optical properties of a porphyrin derivative incorporated in Nafion polymer, Opt. Mat. 27(2005) 1606-1609.